\documentstyle[epsfig,prl,aps,tighten,twocolumn]{revtex}
\newcommand{\be}{\begin{equation}}
\newcommand{\ee}{\end{equation}}
\newcommand{\bea}{\begin{eqnarray}}
\newcommand{\eea}{\end{eqnarray}}
\newcommand{\ba}{\begin{array}}
\newcommand{\ea}{\end{array}}
\newcommand{\nn}{\nonumber}
\newcommand{\E}{{\cal E}}

\newcommand{\eps}{\epsilon}

\begin{document}
\draft
\title{\bf Localization in a random phase-conjugating medium}

\author{M. Blaauboer} 

\address{
Department of Physics, Harvard University, Cambridge, Massachusetts 02138}
\date{\today}
\maketitle

\begin{abstract}
We theoretically study reflection and transmission of light in a one-dimensional
disordered phase-conjugating medium. Using an invariant imbedding approach
a Fokker-Planck equation for the distribution of the probe light reflectance and  
expressions for the average probabilities of reflection and transmission
are derived. A new crossover length scale for localization of light is found,
which depends on the competition between phase conjugation and disorder.
For weak disorder, our analytical results are in good agreement with numerical simulations.
\end{abstract}

\begin{figure}
\centerline{\epsfig{figure=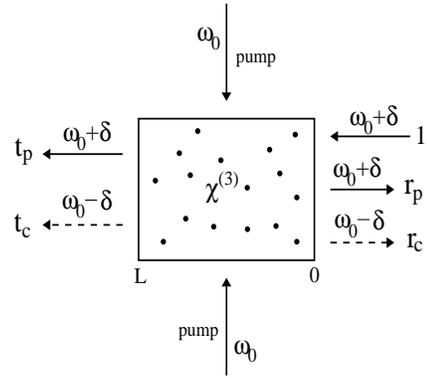,width=5.5cm,height=5cm}}
\vspace*{5mm}
\caption{A 1D disordered phase-conjugating medium, as described in the text. 
Solid (dashed) horizontal lines denote probe (conjugate) beams.}
\label{fig:disPCM}
\end{figure}

Over the last two decades scattering of light from random optical media 
has received a lot of attention\cite{shen90}. In passive 
random media many interesting multiple-scattering effects were discovered, 
such as enhanced backscattering of light\cite{kuga84}, intensity correlations 
in reflected and transmitted waves\cite{garc98} and Anderson 
localization\cite{lage86}. Also absorbing or amplifying random optical 
media have been investigated. 
In the latter, the combination of coherent amplification and confinement
by Anderson localization leads to amplified spontaneous emission and laser 
action without using mirrors, which have been observed in laser 
dyes\cite{lawa94} and semiconductor powders\cite{cao99} resp.
These being linear random media, it is interesting to ask what happens
in a {\it nonlinear} active random medium, such as a disordered 
phase-conjugating medium (PCM). A PCM consists of a nonlinear optical medium with a large 
third-order susceptibility $\chi^{(3)}$, see Fig.~\ref{fig:disPCM}. 
The medium is pumped by two intense counterpropagating 
laser beams of frequency $\omega_{0}$.
When a probe beam of frequency $\omega_{0} + \delta$ is 
incident on the material, a fourth beam will be generated 
due to the nonlinear polarization of the medium. This conjugate 
wave has frequency $\omega_{0} - \delta$ and travels with the reversed phase in the 
opposite direction as the probe beam\cite{fish83}.
The medium thus acts as a "phase-conjugating mirror". Depending on
the characteristics of the PCM, the reflected beam is either stronger
or weaker than the incoming one, while the transmitted probe beam is always
amplified\cite{refl}. It has been shown that phase conjugation also 
occurs in {\it disordered} $\chi^{(3)}$-media\cite{krav90}. 
This raises several interesting questions with respect to reflection 
and transmission of light at such a disordered medium: (1)
how are the amplifying properties of a transparent PCM  
affected in the presence of disorder? (2) What are the 
fundamental similarities and differences between a nonlinear random 
phase-conjugating
medium and a linear amplifying or absorbing random medium? (3) Is there 
a regime 
in which Anderson localization occurs, and what are the requirements 
to observe this?
These questions and their answers form the subject of this paper. 

Our starting point is the wave equation describing a 
one-dimensional (1D) disordered PCM\cite{lens90}
\be
\left( \begin{array}{cc}
\frac{\partial^2}{\partial x^2} + 
k_{p}^{0^{2}}(1 + \eps(x))  & \gamma \vspace{0.3cm} \\
\gamma^{*} &  \frac{\partial^2}{\partial x^2}
+ k_{c}^{0^{2}}(1 + \eps(x)) \end{array} \right)
\Psi(x)
= 0. 
\label{eq:SEFL}
\ee 
Here $k_{p,c}^{0} \equiv (\omega_{0} \pm \delta)/c$ and $\Psi(x) \equiv 
({\cal E}_{p}(x),{\cal E}_{c}^{*}(x))$, with ${\cal E}_{p}(x)$ and 
${\cal E}_{c}^{*}(x)$ the slowly-varying amplitudes of the 
probe and conjugate electric fields respectively. 
The off-diagonal parameter $\gamma \equiv \gamma_{0} e^{i\phi} 
= \frac{6\omega_{0}^2}{\epsilon_{0} c^2}
\chi^{(3)} {\cal E}_{1} {\cal E}_{2}$ is the pumping-induced coupling 
strength between the probe and conjugate waves in the PCM,
with ${\cal E}_{1}$, ${\cal E}_{2}$ the electric field amplitudes of 
the two pump beams. The disorder is modeled by a randomly fluctuating
part $\eps(x)$ in the relative dielectric constant\cite{pump}.
In order to calculate the reflection and transmission coefficients
$r_{p}$, $r_{c}$, $t_{p}$ and $t_{c}$, we use an invariant imbedding 
approach\cite{bell76,ramm87}. Following 
Ref.~\cite{ramm87} we obtain the evolution equations for the probe 
and conjugate waves in the medium
\be
\frac{\partial \E_{p,(c)}^{(*)}(x)}{\partial L} = i k_{0} + i A(L) + B(L) 
+ \frac{i k_{p}^{0}}{2} \eps(L) \E_{p,(c)}^{(*)}(x),
\ee
with
\bea
A(L) & \equiv & \frac{\beta \delta \sqrt{\delta^2 + \gamma_{0}^2}}{\delta^2 + 
\gamma_{0}^2 \cos^2(\beta L)} \\
B(L) & \equiv & \frac{\beta \gamma_{0}^2 \sin(\beta L) \cos(\beta L)}{\delta^2 + 
\gamma_{0}^2 \cos^2(\beta L)},
\eea
$k_{0} \equiv \omega_{0}/c$ and $\beta \equiv \sqrt{\delta^2 + \gamma_{0}^2}/c$. 
Using the boundary conditions from Fig.~\ref{fig:disPCM} at 
$x=0$ and $x=L$ then yields
\begin{mathletters}
\bea
\frac{d r_{p}}{dL} & = & \left[ ik_{0} + i A(L) + B(L) + \frac{i k_{p}^{0}}{2} \eps(L)
(1 + r_{p}) \right] (1 + r_{p}) \nn \\
& & - i k_{p}^{0} ( 1 - r_{p}), \hspace*{2.5cm} r_{p}(0) = 0 
\label{eq:rp} \\
\frac{d r_{c}}{dL} & = & \left[ ik_{0} + i A(L) + B(L) + \frac{i k_{p}^{0}}{2} \eps(L)
(1 + r_{p}) \right] r_{c} \nn \\
& & - i k_{c}^{0} r_{c} - i \frac{\gamma_{0}}{c}, \hspace*{2.5cm} r_{c}(0) = 0 
\label{eq:rc} \\
\frac{d t_{p}}{dL} & = & \left[ ik_{0} + i A(L) + B(L) + \frac{i k_{p}^{0}}{2} \eps(L)
(1 + r_{p}) \right] t_{p}, \nn \\
& & \hspace*{4.5cm} t_{p}(0) = 1 
\label{eq:tp} \\
\frac{d t_{c}}{dL} & = & \left[ ik_{0} + i A(L) + B(L) + \frac{i k_{p}^{0}}{2} \eps(L)
(1 + r_{p}) \right] t_{c}, \nn \\
& & \hspace*{4.5cm} t_{c}(0) = 0.
\label{eq:tc}
\eea
\label{eq:coef}
\end{mathletters}
In the absence of phase conjugation, for $\gamma_{0} = 0$, equations~(\ref{eq:rp}) and 
(\ref{eq:tp}) reduce to the well-known imbedding equations for a linear random 
medium\cite{ramm87}, and $r_{c}= t_{c}=0$. In the absence of disorder,  
equations~(\ref{eq:rc}) and (\ref{eq:tp}) reduce to the evolution 
equations for $r_{c}$ and $t_{p}$ in a transparent PCM, and $r_{p}=t_{c}=0$.
Equations~(\ref{eq:coef}) satisfy the energy conservation law $R_{p} + T_{p} - 
R_{c} - T_{c} = 1$, with $R_{p} \equiv |r_{p}|^2$ the probe reflectance etc.
They form the basis of all our results here. We first derive 
a Fokker-Planck (FP) equation for the probability distribution of $R_{p}$.
We set $r_{p} \equiv R_{p} e^{i \Theta_{p}}$, substitute this into 
(\ref{eq:rp}), subsequently into the Liouville equation 
$\frac{\partial Q}{\partial L} = - \frac{\partial}{\partial R_{p}} \left[ Q 
\frac{d R_{p}}{d L} 
\right] - \frac{\partial}{\partial \Theta_{p}} \left[ Q \frac{d \Theta_{p}}{d L} 
\right]$, where $Q(R_{p},\Theta_{p})$ is the density of points 
($R_{p}$,$\Theta_{p}$) in phase space, and average over the disorder. 
Assuming a gaussian distribution for $\eps(L)$, with $\langle \eps(L) \rangle=0$
and $\langle \eps(L) \eps(L^{'}) \rangle = g \delta (L - L^{'})$, where pointed brackets
denote an average over disorder, yields
\bea
\frac{\partial W}{\partial l} & = R_{p}(1 - R_{p})^2\,  \frac{\partial^2 W}{\partial R_{p}^2}
+ (1 - 2 (3 + M(L)) R_{p} + 5 R_{p}^2)  \nn \\ & \frac{\partial W}{\partial R_{p}} 
- 2 ( 1 +  M(L) - 2 R_{p}) W,\   W(0) = \delta (R_{p}).
\label{eq:FP}
\eea
Here $W \equiv \langle Q \rangle$, $l \equiv L/\xi_{0}$,
$M(L) \equiv \xi_{0} B(L)$ and $\xi_{0}^{-1} \equiv \frac{1}{2} g k_{p}^{0^{2}}$,  
the inverse localization length in the absence of phase conjugation.
In deriving (\ref{eq:FP}) we have neglected angular variations of $W$, the
random-phase approximation (RPA), which applies to the situation of weak disorder
when $\xi_{0} \gg 1/\beta$. Equation~(\ref{eq:FP}) is of the same form as 
the equation for the probability distribution of the reflectance at a
linear active random medium\cite{prad94}, with the important difference that in the latter 
case M(L) is an $L$-independent constant, proportional to the imaginary part of the
dielectric constant. Using this analogy, our phase-conjugating medium alternates 
between a linear amplifying (for $M(L) < 0 $) and linear absorbing (for $M(L) > 0$) 
random medium. 
For $M(L)=0$ the well-known FP equation for a passive random medium 
$\frac{\partial W}{\partial l} = \frac{\partial}{\partial R_{p}}\left[ R_{p} 
\frac{\partial}{\partial R_{p}} (1 - R_{p})^2 W \right]$ \cite{abri81} is retrieved.

Multiplying both sides of~(\ref{eq:FP}) by $R_{p}^{n}$ and integrating by parts
leads to a recursion relation for the moments of the probe reflectance,
\be
\frac{d}{d l} \langle R_{p}^{n} \rangle = n^2 \langle R_{p}^{n+1} 
\rangle - 2n (n - M(L)) \langle R_{p}^{n} \rangle + n^2 \langle R_{p}^{n-1} \rangle.
\label{eq:momrp}
\ee
For $n=1$ and setting $\langle R_{p}^{2} \rangle \approx \langle R_{p}\rangle$\cite{bimodaal},
integration of~(\ref{eq:momrp}) yields for the average probe reflectance
\bea
\langle R_{p} \rangle & = & \left[ C + \alpha^2 \gamma_{0}^2 \cos(2\beta L) + 
2 \alpha \beta \gamma_{0}^2 \sin(2 \beta L) - (C + \right. \nn \\  && \left.
\alpha^2 \gamma_{0}^2) e^{- \alpha L}
\right] / [ C + (4 \beta^2 + \alpha^2) \gamma_{0}^2 \cos(2 \beta L)],
\label{eq:probe}
\eea
with $C \equiv (4 \beta^2 + \alpha^2)(2\delta^2 + \gamma_{0}^2)$ and
$\alpha \equiv 1/\xi_{0}$. In the absence of phase conjugation, this reduces to 
$\langle R_{p}\rangle = 1 - e^{-L/\xi_{0}}$\cite{kuma85} and in the absence of
disorder $\langle R_{p}\rangle = 0$, as for a transparent PCM. Using equations~(\ref{eq:rp})
and (\ref{eq:rc}), one can directly obtain an evolution equation for the 
average $Z_{n,m} \equiv \langle R_{p}^{n} R_{c}^{m} \rangle$,
which is given by
\bea
\frac{d}{d l} Z_{n,m} & = & n (m+n) 
Z_{n+1,m} - [2n^2 + m(n+1)] Z_{n,m} + \nn \\ 
& n^2 & Z_{n-1,m} + 2(m+n) M(L) Z_{n,m} + 2m M(L),
\label{eq:momrprc}
\eea
and equivalent to~(\ref{eq:momrp}) for $m=0$. 
Solving~(\ref{eq:momrprc}) for the conjugate reflectance yields
\be
\langle R_{c} \rangle = -1 + \frac{\delta^2 + \gamma_{0}^2}{\delta^2 + \gamma_{0}^2
\cos^2(\beta L)}\ e^{-\alpha L} + \langle R_{p} \rangle.
\label{eq:rcav}
\ee
Similarly, one obtains for the probe transmittance from (\ref{eq:rp}) and 
(\ref{eq:tp})\cite{general} 
\be
\langle T_{p} \rangle = 
\frac{\delta^2 + \gamma_{0}^2}{\delta^2 + \gamma_{0}^2 \cos^2(\beta L)}\ 
e^{-\alpha L}. 
\label{eq:probetrans}
\ee
The conjugate transmittance is then given by $\langle R_{c} \rangle =0$, 
through the conservation law $\langle R_{p} \rangle + \langle T_{p} \rangle - 
\langle R_{c} \rangle - \langle T_{c} \rangle = 1$.

\begin{figure}
\centerline{\epsfig{figure=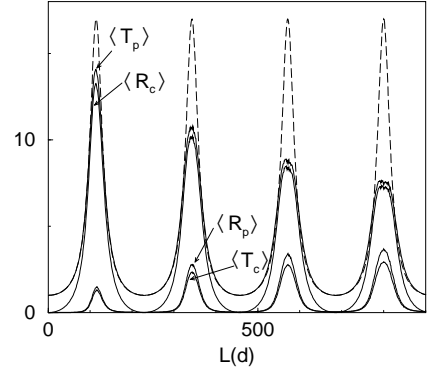,width=5.5cm,height=5cm}}
\vspace*{5mm}
\caption[]{Probe and conjugate reflectance and transmittance (averaged over
10000 realizations of the disorder) at a disordered phase-conjugating medium as a function
of the length $L$ (in units of the lattice spacing $d$) of the medium. 
The dashed curve denotes $T_{p}$ in the absence of disorder. 
Parameters used are $\delta = 10^{10} s^{-1}$, $\gamma_{0} = 4 \cdot 10^{10} s^{-1}$
and $\eps(x) \in [-0.05,0.05]$, corresponding to a localization length $\xi_{0} = 1.2$ m.
}
\label{fig:plot1}
\end{figure}

\begin{figure}
\centerline{\epsfig{figure=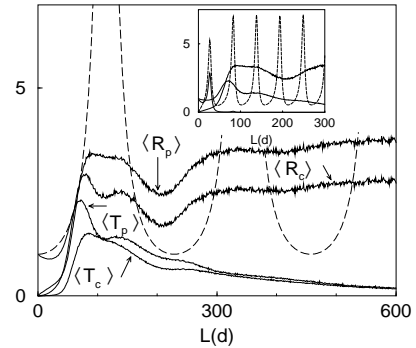,width=5.5cm,height=5cm}}
\vspace*{5mm}
\caption[]{Same as Fig.~\ref{fig:plot1} but now for larger disorder, 
$\eps(x) \in [-0.5,0.5]$, corresponding to a localization length $\xi_{0} = 12$\, mm.
The inset compares the analytic results (\ref{eq:probe}) for $\langle R_{p} \rangle$
(dashed curve) and (\ref{eq:probetrans}) for $\langle T_{p} \rangle$ 
(thick solid curve) with the numerical ones.
}
\label{fig:plot2}
\end{figure}

\begin{figure}
\centerline{\epsfig{figure=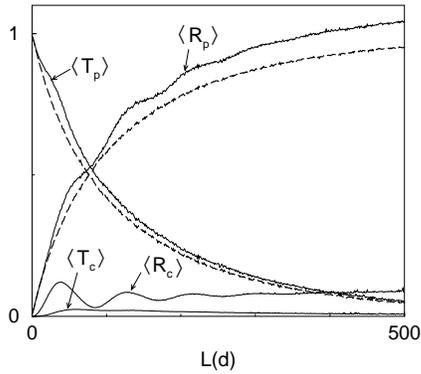,width=5.5cm,height=5cm}}
\vspace*{5mm}
\caption[]{Same as Fig.~\ref{fig:plot2}, but for $\delta = 10^{11} s^{-1}$.
The thick dotted curves denote $\langle R_{p} \rangle$ and $\langle T_{p} \rangle$ 
in the absence of phase conjugation. 
}
\label{fig:plot3}
\end{figure}

In order to test these analytical predictions we have carried out 
numerical simulations. Using a transfer matrix method\cite{datt93},
equations~(\ref{eq:SEFL}) are discretized on a 1D lattice
with lattice constant $d$, into which disorder is introduced by 
letting $\eps(x)$ randomly fluctuate
from site to site. Figures~(\ref{fig:plot1})-(\ref{fig:plot3}) show the
probe and conjugate reflectance and transmittance as a function of the length
$L$ of the medium for various values of the detuning $\delta$ and disorder.
In all cases we took $d= 10^{-4}$ m and $\omega_{0} = 10^{15} s^{-1}$
and typical PCM parameters.

Fig.~\ref{fig:plot1} shows how the periodic behavior of $\langle R_{c} \rangle$ and 
$\langle T_{p} \rangle$ which is characteristic of a transparent PCM becomes "modulated"
by an exponentially decaying envelope in the presence of weak disorder.
Simultaneously, and with the same periodicity, some probe light is now
reflected and some conjugate light transmitted, due to normal reflections in the 
disordered medium. When the amount of disorder is increased, the oscillatory
behavior of the reflectances and transmittances is less and becomes suppressed
for large $L$, see Fig.~\ref{fig:plot2}. The reflected 
probe and conjugate intensities then both saturate, with 
$\lim_{L \rightarrow \infty} 
\langle R_{c} \rangle =  \lim_{L \rightarrow \infty} \langle R_{p} 
\rangle -1$, and
$\langle T_{p} \rangle$ and $\langle T_{c} \rangle$ decay to zero (localization).
For a transparent PCM the conservation law $T_{p} - R_{c} = 1$ applies, i.e. 
for each pump photon scattered into the forward (probe) beam in the 
medium, a photon from the other pump is 
scattered into the backward (phase-conjugate) beam. In the localization regime of 
Fig.~\ref{fig:plot2}, on the other hand, the conservation law $\langle R_{p} \rangle - 
\langle R_{c} \rangle = 1$ applies (cf. Eq.~(\ref{eq:rcav})). Hence $\langle T_{p}
\rangle$ has exchanged roles with $R_{p}$ due to disorder: all pump 
photons which are absorbed into probe
and conjugate beams are now reflected and despite amplification, 
transmitted intensities are suppressed. This suppression has also 
been found in linear amplifying random media\cite{paas96,been96}.
The saturation of $\langle R_{c} \rangle$ suggests that the phase-conjugate 
reflected beam arises in the region into which the
probe beam penetrates and that amplification takes mostly place 
within a localization length
of the point of incidence. The behavior of the transmitted intensities 
with increasing length of the medium is determined by two competing 
effects: on the one hand, enhancement occurs due to increased probability of multiple 
reflections. On the other hand, less light is transmitted due to 
increased probability of retroreflection
of the incoming probe light. 
For small $L$, the latter effect dominates $\langle T_{p} \rangle$ in Fig.~\ref{fig:plot2}.
As $L$ increases, the increasing amplification of probe light due 
to multiple scattering takes over, which leads to exponential increase and 
a maximum in $\langle T_{p} \rangle$. For again larger $L$, most of the probe light is 
reflected, and $\langle T_{p} \rangle$ decreases exponentially to zero, as in a normal 
disordered medium\cite{cross}. The crossover length scale $L_{c}$ between 
exponential increase and decrease is given by the solution 
of $\delta^2 + \gamma_{0}^2 \cos^2(\beta L) = 
2 \gamma_{0}^2 \beta \xi_{0} \cos (\beta L) \sin (\beta L)$, which for 
$\delta \ll \gamma_{0}$ becomes
\be
L_{c} \approx \frac{c}{\gamma_{0}} \left( \frac{\pi}{2} -  \arctan \left( 
\frac{c}{2\gamma_{0} \xi_{0}} \right) \right).
\label{eq:crossover}
\ee
In the opposite limit of $\delta \gg \gamma_{0}$ phase-conjugate reflection
is weak (maximum value of $R_{c} = 0.16$) and we retrieve exponential localization, 
see Fig.~\ref{fig:plot3}. Randomness now dominates over phase conjugation 
and has almost washed out the oscillatory 
behavior of $\langle R_{p} \rangle$ and $\langle T_{p} \rangle$.

Comparing the numerical results with the analytic ones from (\ref{eq:probe}),
(\ref{eq:rcav}) and (\ref{eq:probetrans}) we find good agreement
(deviations $<$ 5 \%) for weak disorder as in Fig.~\ref{fig:plot1}, and
for stronger disorder and weak phase conjugation as in Fig.~\ref{fig:plot3}. 
In the intermediate regime, for $\xi_{0} > 1/\beta 
\approx c/\gamma_{0}$ results differ considerably, 
see inset in Fig.~\ref{fig:plot2}. There the RPA and the assumption 
$\langle R_{p}^2 \rangle \approx \langle R_{p} \rangle$ are not 
valid and a different approach is needed. 
In conclusion, we have studied reflection and transmission of light at a 1D
disordered phase-conjugating medium in the limit of small disorder, for 
$\beta \xi_{0} \gg 1$. The predicted behavior of reflectances and transmittances
arising from the interplay between amplification and Anderson localization displays
similar features as that in a linear disordered amplifying medium.
The main difference is the coupling of two waves in the PCM, which 
leads to additional interference effects.
In future work we intend to: (1) investigate the strong disorder regime. 
There the reflection
of the pump beams cannot be neglected, and a full nonlinear analysis is required.
(2) Study the distribution of reflection and transmission eigenvalues and the 
statistical fluctuations in reflectance and transmittance for a multimode 
2D or 3D disordered phase-conjugating medium\cite{3D}. This is relevant
to experiments, which mostly employ 3D PCM's\cite{lanz96}, and interesting in 
the context of random lasers: in a linear disordered amplifying medium 
the average reflectance 
becomes infinitely large with increasing amplification, upon approaching 
threshold\cite{been96}. It would be interesting to investigate whether 
something similar occurs in a disordered PCM, this being a "naturally" 
amplifying medium and feasible candidate for nonlinear random lasing.

The author gratefully acknowledges stimulating discussions
with D. Lenstra. This work was supported by
the Netherlands Organisation for Scientific Research (NWO).


\begin{references}
\bibitem{refl}
The conjugate reflectance $R_{c}$ (defined
as the reflected conjugate intensity divided by the incoming probe intensity)
and probe transmittance $T_{p}$ are given by \cite{fish83}
$R_{c} = \sin^2(\beta L)/(\cos^2(\beta L) + ( \delta^2 /\gamma_{0}^2 ))$ and 
$T_{p} = (1 + ( \delta^2/\gamma_{0}^2))/(\cos^2(\beta L) +
( \delta^2/\gamma_{0}^2))$, with $\beta \equiv \sqrt{\delta^2 + \gamma_{0}^2}/c$.
\bibitem{pump}
In the presence of disorder some of the pump light will be retroreflected. 
But since the pumps are typically 3 orders of magnitude stronger 
than the probe we neglect this reflection for weak disorder. Also, as long as
probe and conjugate reflected and transmitted intensities are sufficiently small, 
pump depletion can be neglected and the pumps may 
be taken constant in the medium.
\bibitem{bimodaal}
This is appropriate if $R_{p} \approx 0$ or $R_{p} \approx 1$,
in the regime of either weak disorder or combination of strong disorder 
and weak phase conjugation.
\bibitem{3D}
The 1D invariant imbedding approach used here can straightforwardly 
be extended to 2D and 3D disordered PCM's, by decomposing the system
into one longitudinal dimension (direction of propagation) and
one or two transverse dimension(s) (transverse modes).
\end{references}
\end{document}